\documentclass[osajnl,twocolumn,showpacs,superscriptaddress,10pt]{revtex4-1} 
\usepackage{amsmath,amssymb,graphicx}

\usepackage{mathtools}
\usepackage{xfrac}

\begin{document}

\title{Dual-pumped degenerate Kerr oscillator in a silicon nitride microresonator }

\author{Yoshitomo Okawachi}\email{Corresponding author: y.okawachi@columbia.edu}
\affiliation{School of Applied and Engineering Physics, Cornell University, Ithaca, NY 14853, USA}
\affiliation{Department of Applied Physics and Applied Mathematics, Columbia University, New York, NY 10027}

\author{Mengjie Yu}
\affiliation{School of Applied and Engineering Physics, Cornell University, Ithaca, NY 14853, USA}
\affiliation{Department of Applied Physics and Applied Mathematics, Columbia University, New York, NY 10027}

\author{Kevin Luke}
\affiliation{School of Electrical and Computer Engineering, Cornell University, Ithaca, NY 14853, USA}

\author{Daniel O. Carvalho}
\affiliation{School of Electrical and Computer Engineering, Cornell University, Ithaca, NY 14853, USA}
\affiliation{current address: S\~{a}o Paulo State University (UNESP), S\~{a}o Jo\~{a}o da Boa Vista, Brazil}

\author{Sven Ramelow}
\affiliation{School of Applied and Engineering Physics, Cornell University, Ithaca, NY 14853, USA}
\affiliation{Faculty of Physics, University of Vienna, 1090 Vienna, Austria}

\author{Alessandro Farsi}
\affiliation{School of Applied and Engineering Physics, Cornell University, Ithaca, NY 14853, USA}

\author{Michal Lipson}
\affiliation{School of Electrical and Computer Engineering, Cornell University, Ithaca, NY 14853, USA}
\affiliation{Department of Electrical Engineering, Columbia University, New York, NY 10027}
\affiliation{Kavli Institute at Cornell for Nanoscale Science, Ithaca, NY 14853, USA}

\author{Alexander L. Gaeta}
\affiliation{School of Applied and Engineering Physics, Cornell University, Ithaca, NY 14853, USA}
\affiliation{Department of Applied Physics and Applied Mathematics, Columbia University, New York, NY 10027}
\affiliation{Kavli Institute at Cornell for Nanoscale Science, Ithaca, NY 14853, USA}

\begin{abstract}We demonstrate a degenerate parametric oscillator in a silicon-nitride microresonator. We use two frequency-detuned pump waves to perform parametric four-wave mixing and operate in the normal group-velocity-dispersion regime to produce signal and idler fields that are frequency degenerate. Our theoretical modeling shows that this regime enables generation of bimodal phase states, analogous to the $\chi^{(2)}$-based degenerate OPO. Our system offers potential for realization of CMOS-chip-based coherent optical computing and an all-optical quantum random number generator.
\end{abstract}

\ocis{(190.4380) Four-wave mixing; (190.4970) Parametric oscillators and amplifier; (190.4390) Integrated optics.}

\maketitle 


Researchers recently proposed using degenerate optical parametric oscillators (OPO's) to realize a novel form of coherent optical computing by simulating the quantum Ising model \cite{Wang,Marandi14,Fabre}. Such an approach offers the possibility of solving non-deterministic polynomial time (NP)-hard computation problems \cite{Barahona}. A first proof-of-principle demonstration based on a degenerate 2nd-order nonlinearity $\chi^{(2)}$ OPO using bulk optics offers the promise that this new paradigm could be realized  \cite{Fabre,Marandi12}. The approach is based on the non-equilibrium phase transition that occurs at threshold for optical parametric oscillation, in which the generated field with respect to the pump field locks to one of two possible states offset by $\pi$. Below threshold, the signal field is described by a squeezed vacuum state, where the strength of squeezing depends on the ratio of the intracavity power to the threshold power \cite{Ou,Collett}. The vacuum fluctuations of the out-of-phase quadrature are suppressed, while those of the in-phase quadrature are enhanced. It is along the in-phase quadrature that, at threshold, the system bifurcates into two possible states with opposite phases. By utilizing a network of coupled OPO's, more complex phase-locked states can be achieved which encode the ground-state of the quantum Ising model \cite{Wang}, where the coupling strengths of its spins are equivalent to the coupling between the OPO's. It has been suggested that such an OPO system offers potential for a simulation device that can outperform any other classical computation \cite{Haribara}. Recently, a 4-OPO system based on the $\chi^{(2)}$ nonlinearity has been demonstrated \cite{Marandi14}. Alternatively, the bi-phase state output can be used to realize a quantum random number generator \cite{Marandi12} that compared to other implementations requires only classical signal detection (no single photon detection), is intrinsically unbiased, and thus does not need any post-procession of the raw output, which has applications in cryptography \cite{Ekert}, Monte Carlo simulations \cite{Alspector}, and quantitative finance \cite{Banks}. The scalability and robustness of these coupled OPO's will be crucial for these applications. 

\begin{figure}[t]
\centerline{\includegraphics[width=\linewidth]{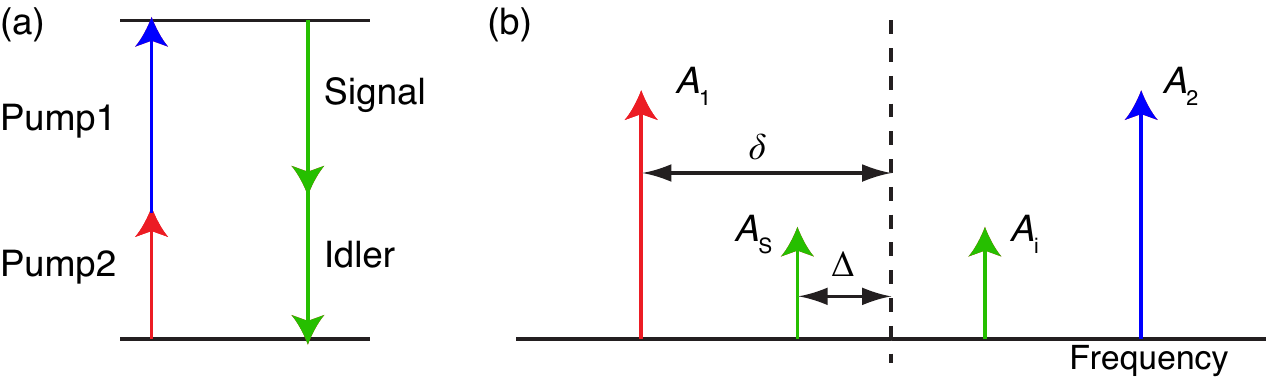}}
\caption{(a) Four-wave mixing energy level diagram in which two pump pumps are converted to two frequency-degenerate photons.  (b) Scheme for our $\chi^{(3)}$ OPO in Si$_3$N$_4$ microresonator, where $A_1$ and $A_2$ are the pump fields, $A_s$ and $A_i$ are the generated signal and idler. $\delta$ is the frequency offset between the pumps and the center frequency, $\Delta$ is the frequency offset from degeneracy between the signal/idler and the center frequency.}
\label{Fig1}
\end{figure}

An alternative approach suggested in \cite{Marandi14} is to utilize a $\chi^{(3)}$ nonlinearity to realize an analogous degenerate OPO with dynamics that would mirror the $\chi^{(2)}$ system. The silicon-based photonics platform appears to be highly suited for this application since it allows for a robust, compact, integrated system that offers feasibility for scaling and stability. In recent years, there have been various nonlinear optics demonstrations in microresonators \cite{Heebner,Turner08,Kippenberg,LevyOE,Levy,Okawachi,Herr12,Peccianti,Jung,Savchenkov, Papp,Liu,Herr,Hausmann,Huang,Kalchmair}. In particular, there has been significant development in frequency comb generation through optical parametric oscillation in high-\emph{Q} microresonators \cite{Kippenberg,Levy,Okawachi,Herr12,Jung,Savchenkov,Papp,Liu,Herr,Hausmann,Huang} that utilizes parametric four-wave mixing (FWM) interactions based on the third-order nonlinearity $\chi^{(3)}$. The silicon nitride (Si$_3$N$_4$) platform offers potential as a platform for coupled degenerate OPO's due to its CMOS compatibility, low linear and nonlinear losses at telecommunication wavelengths, and high nonlinearity and dispersion engineering \cite{Moss}. Recently, there have been demonstrations of inverse FWM using optical fibers in the normal group-velocity dispersion (GVD) regime \cite{Turitsyn}, and Inagaki, \emph{et al.} \cite{Inagaki} has shown in initial experiments that a dual-pumped $\chi^{(3)}$ fiber OPO's multiplexed in the time domain can be used to simulated a 1-D Ising model. 

\begin{figure}[tp]
\centerline{\includegraphics[width=\linewidth]{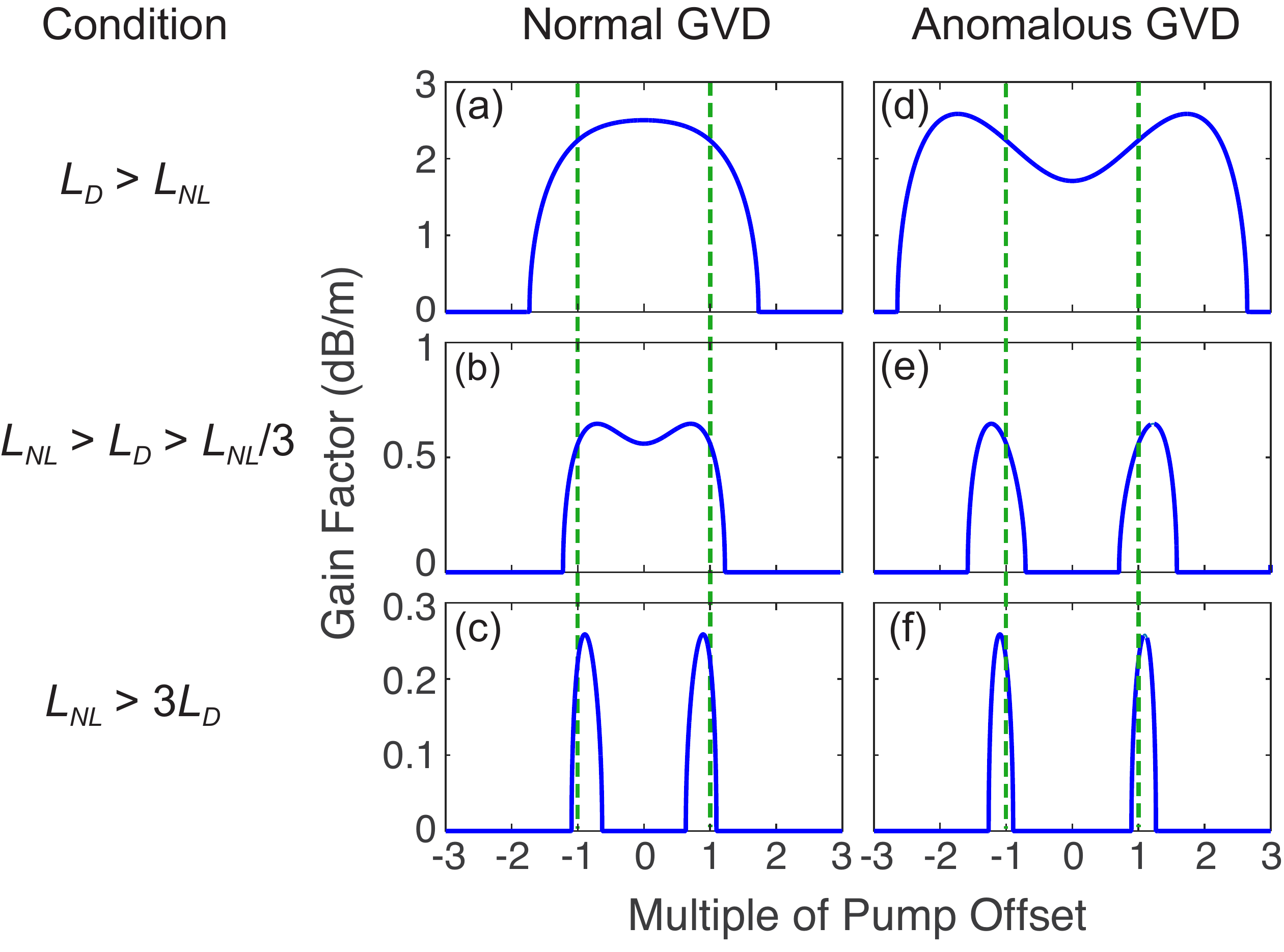}}
\caption{Parametric gain profile assuming two undepleted nondegenerate pumps. Dashed lines indicate the position of the pumps. Pump separation is 14 nm and the center wavelength between the pumps is 1550 nm. (a)-(c) show the gain profiles in normal GVD ($\beta_2 = 193 $ ps$^2$/km) for the ratio between $L_D$ and $L_{NL}$ of 2, 0.5 and 0.2, respectively, The nonlinear coefficient ($\gamma = 0.94 $ W$^{-1}$m$^{-1}$) is based on the 690$\times$1300 nm cross-section Si$_3$N$_4$ waveguide used in our experiment. (d)-(f) show the gain profiles in anomalous GVD ($\beta_2 = -193$ ps$^2$/km) for the same ratios. The ratio between $L_D$ and $L_{NL}$ is adjusted by changing the pump power. The effect of higher-order dispersion terms is negligible because of the small frequency separation between the pumps as well as the waveguide dispersion.}
\label{Fig2}
\end{figure}

In this paper, we theoretically and experimentally demonstrate that a dual-pumped, microresonator-based $\chi^{(3)}$ OPO can achieve degenerate oscillation. In our scheme, a Si$_3$N$_4$ microresonator is dual-pumped to generate a frequency-degenerate signal and idler pair through parametric FWM [Fig. \ref{Fig1}(a)]. We perform theoretical analysis to determine the conditions under which degenerate oscillation and the associated random binary phase state is achieved and find that it requires normal GVD using a dual-pump. We show experimentally that frequency-degenerate oscillation is achieved by controlling the frequency offset between the two pumps.   

Our configuration consists of two pump fields denoted by $A_1$ and $A_2$ that are offset by frequency $\pm \delta$ from the degeneracy point [see Fig. \ref{Fig1}(b)]. There has been a number of experimental studies that  have investigated parametric FWM with two pump waves \cite{Radic,Tong,Marhic,Inoue}. Energy conservation implies that the generated signal and idler fields with amplitudes denoted by $A_s$ and $A_i$, respectively, can grow at a frequency offset $\Delta$ from degeneracy for the signal/idler pair. In order for degenerate oscillation to occur, the parametric gain for the signal and idler must peak at $\Delta=0$. We model the nonlinear interaction using the nonlinear Schr\"{o}dinger equation, 

\begin{equation}
\frac{dA}{dz} = -i\frac{\beta_2}{2}\frac{\partial^2A}{\partial \tau^2} +i\gamma |A|^2A,
\label{eq:Schro}
\end{equation}

\noindent where $\beta_2$ is the GVD parameter and $\gamma$ is the nonlinear parameter proportional to $\chi^{(3)}$. We write the total field with respect to the center mode frequency as

\begin{equation}
A(z,\tau) = A_1(z)e^{i\delta \tau}+A_2(z)e^{-i\delta \tau}+A_s(z)e^{i\Delta \tau}+A_i(z)e^{-i\Delta \tau}.
\label{eq:Field}
\end{equation}

\noindent We assume that the pump waves are undepleted and have equal powers $P=|A_1|^2=|A_2|^2$. The solution for the signal/idler pair has an exponential form $e^{gz/2}$, where $g$ is the gain parameter, 

\begin{equation}
g = \sqrt{[6\gamma P-\beta_2(\delta^2-\Delta^2)]\times[2\gamma P+\beta_2(\delta^2-\Delta^2)]}.
\label{eq:Field}
\end{equation}

\noindent Our analysis indicates that with suitable control of the pump power and offset frequency, it is possible to find a condition for achieving parametric gain at modest normal GVD (i.e., $\beta_2>0$), which can be readily achieved by engineering the waveguide cross-section \cite{Turner}. We investigate the parametric gain profile based on the parameters of a 690$\times$1300 nm cross section Si$_3$N$_4$ waveguide pumped with two single-frequency lasers. Figure \ref{Fig2} shows the dependence of the gain on the dispersion length $L_D=1/\delta^2|\beta_2|$ and the nonlinear length $L_{NL}=1/2\gamma P$. We show three different regimes, each with a different ratio between $L_D$ and $L_{NL}$ for both normal and anomalous GVD and find that maximum parametric gain can be achieved at the degenerate signal/idler frequency only when $L_D > L_{NL}$ in the normal GVD regime [Fig. \ref{Fig2}(a)]. Operating in this well-defined normal GVD regime is essential in order to implement the bi-phase state, as this ensures that the degenerate pairs have the lowest threshold for oscillation. 

\begin{figure}[t]
\centerline{\includegraphics[width=\linewidth]{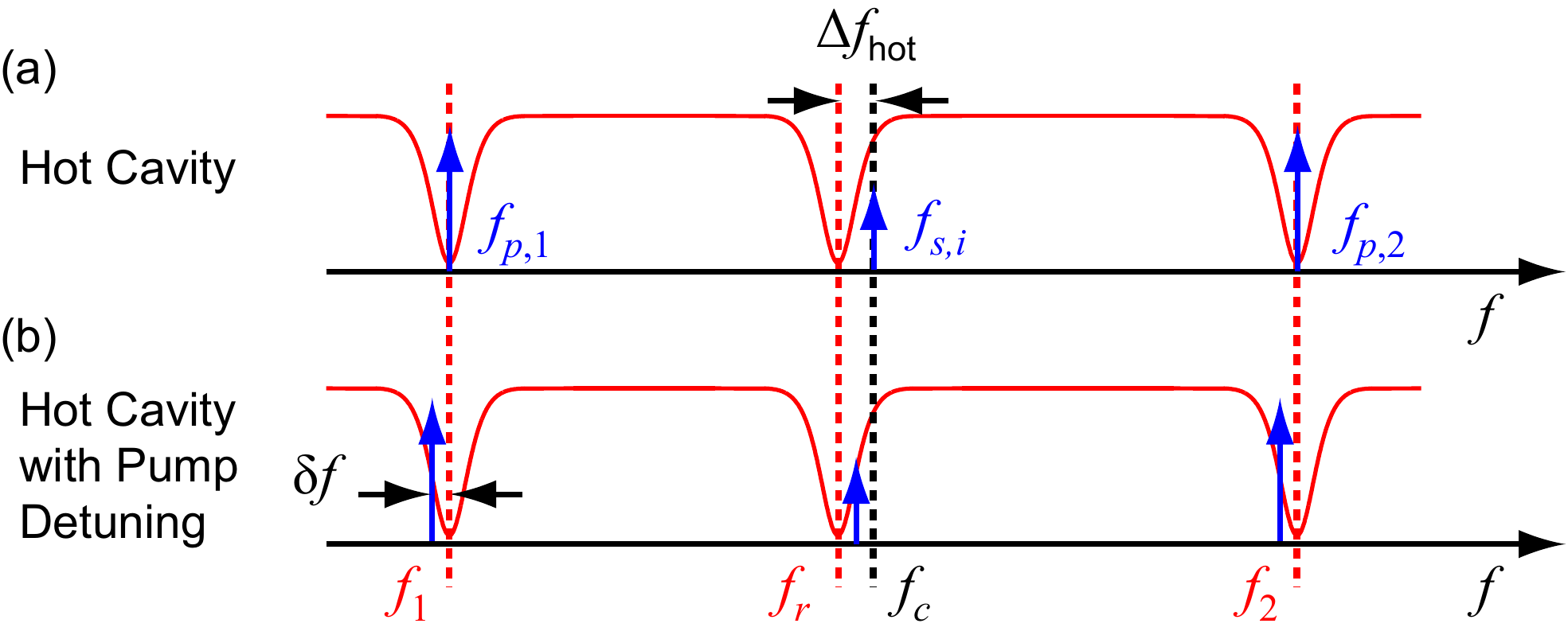}}
\caption{Resonator condition for dual-pump OPO. The relative positions of the resonances are shown for (a) a hot cavity with zero pump detuning and (b) hot cavity with pump detuning. $f_1$, $f_2$, and $f_r$ (red) are the resonance frequencies for pump 1, pump 2, and degenerate signal/idler modes, respectively. $f_{p,1}$, $f_{p,2}$, $f_{s,i}$ (blue) are the two pump frequencies and the generated signal/idler frequency, respectively. $f_c=\frac{f_1+f_2}{2}$, $\Delta f_{\text{hot}}=f_r-f_c$, and the pump detuning $\delta f=f_{p,1}-f_1=f_{p,2}-f_2$.}
\label{Fig3}
\end{figure}

Furthermore, in order to ensure degenerate oscillation in a microresonator cavity, the position of the two pumps and the signal/idler within the cavity resonance is critical. Figure \ref{Fig3} shows the resonator conditions in the normal GVD regime. We define the resonator mode frequencies near the frequencies of pump 1, pump 2, and the degenerate signal/idler frequencies as $f_1$, $f_2$, and $f_r$, respectively (indicated in red in Fig. \ref{Fig3}). The center frequency between the two pump resonance modes $f_c=\frac{f_1+f_2}{2}$ is offset from the signal/idler resonance mode by $\Delta f=f_r-f_c$. This frequency offset is dependent on the dispersion and, for a cold cavity with normal GVD, it is given by $\Delta f_{\text{cold}}=v_g/4\pi L_D$, where $v_g$ is the group velocity. In a hot cavity, the frequency offset is due to both dispersion and the Kerr nonlinearity and is given by $\Delta f_{\text{hot}}=\frac{v_g}{4\pi}(\frac{1}{L_D}-\frac{1}{L_{NL}})$. For pump frequencies of $f_{p,1}$ and $f_{p,2}$ (indicated in blue in Fig. \ref{Fig3}), we define detuning of the pump frequency from the peak of the resonance as $\delta f=f_{p,1}-f_1=f_{p,2}-f_2$. The signal/idler position with respect to the resonance for the zero detuning case, where both pumps are tuned to the peak of the respective resonances, is shown in [Fig. \ref{Fig3}(a)]. As discussed earlier, for the degenerate signal/idler to oscillate, $L_D > L_{NL}$ must be satisfied, which corresponds to a negative $\Delta f_{\text{hot}}$ that could jeopardize the oscillation of the degenerate pair. If $|\Delta f_{\text{hot}}|>\delta \nu/2$, where $\delta \nu$ is the cavity linewidth, the degenerate pair is suppressed. However, we can take advantage of pump detuning to achieve the optimum oscillation conditions in the resonator. First, since $L_{NL}$ is dependent on the total power built-up inside the resonator, pump detuning affects the parametric gain profile required for oscillation. In addition, the pump frequency tuning can be used to control the optimal signal/idler frequency position provided that $|\Delta f_{\text{hot}} - \delta f| <\delta \nu/2$ [Fig. \ref{Fig3}(b)]. This allows for all fields to lie within their respective cavity resonances allowing for degenerate oscillation to occur. In practice, the ultimate detuning range that can be achieved is limited by the thermal effect in the resonator from pumping two different resonances.

\begin{figure}[tp]
\centerline{\includegraphics[width=\linewidth]{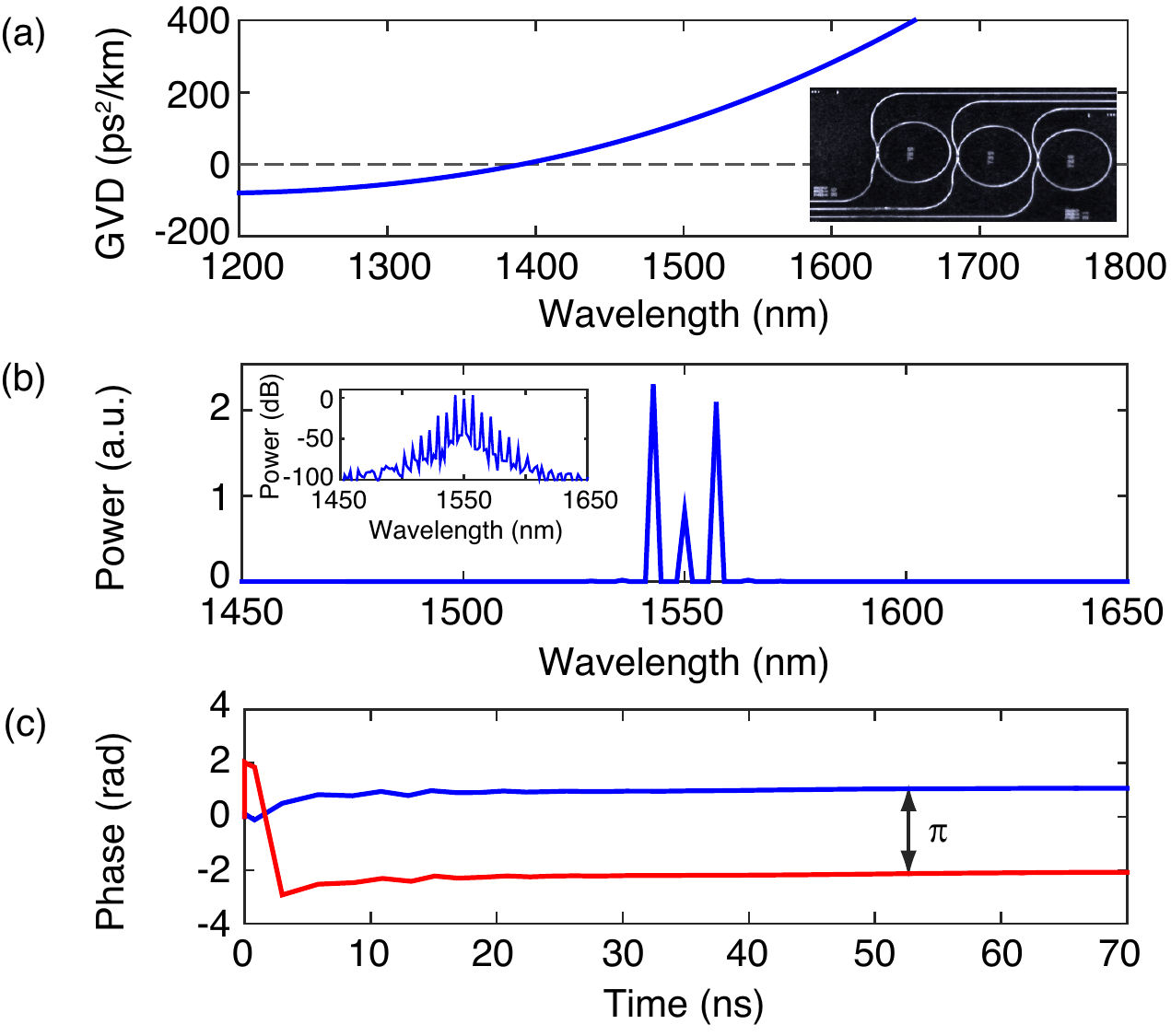}}
\caption{(a) Simulated GVD for the fundamental TM mode of 690$\times$1300 nm cross section Si$_3$N$_4$ microresonator. Inset: Si$_3$N$_4$ microresonators. (b) Simulated intracavity spectrum for degenerate OPO in Si$_3$N$_4$ microresonator. Inset shows the spectrum on a dB scale. The pump offset from the degenerate frequency $\delta= 4\times$FSR. (c) Evolution of the phase of the signal for two realizations. Bi-phase states are achieved with a $\pi$ phase offset. }
\label{Fig4}
\end{figure}

\begin{figure}[t]
\centerline{\includegraphics[width=\linewidth]{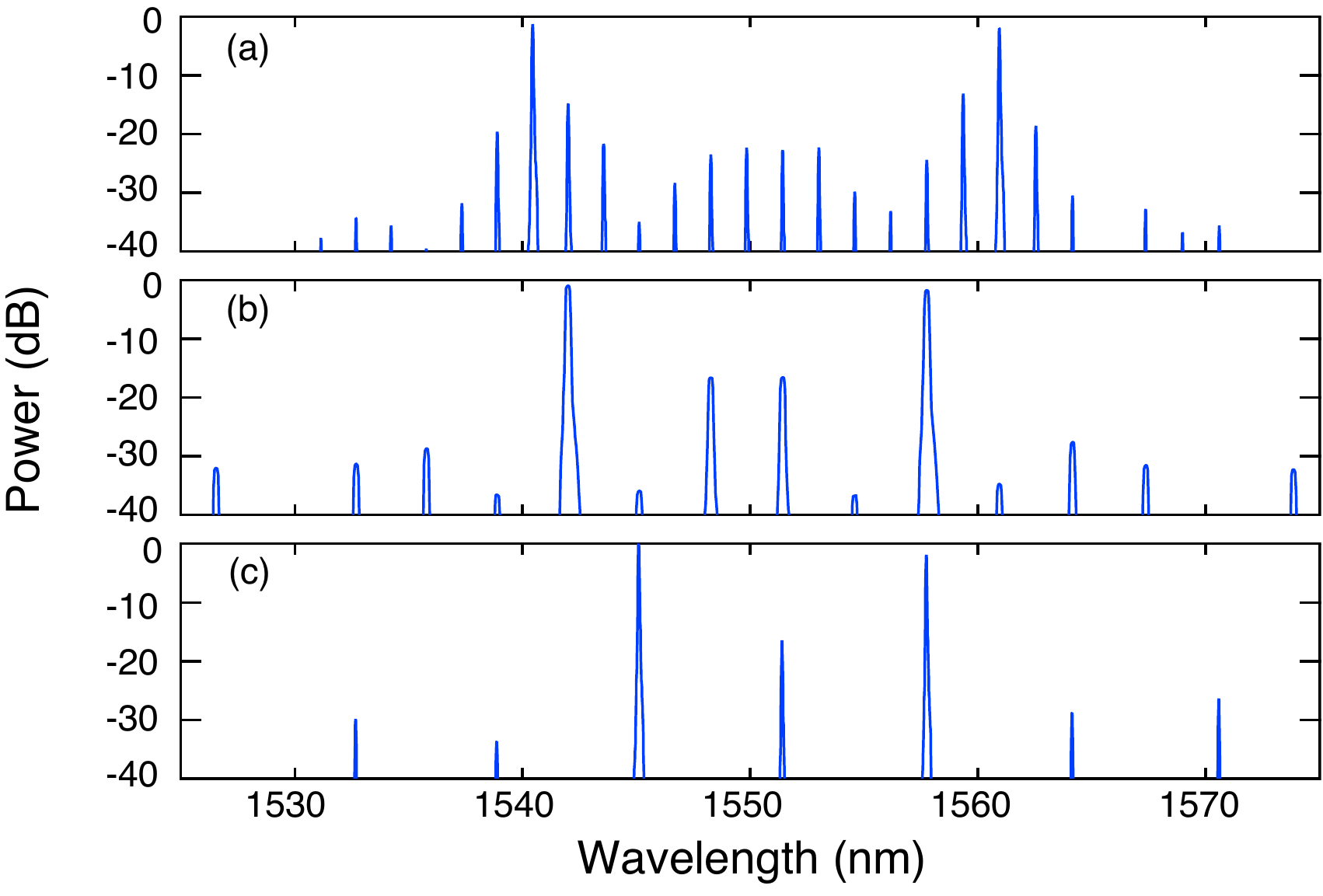}}
\caption{OPO spectra from two pumps for pump offsets $\delta$ of (a) 6.5$\times$FSR, (b) 5$\times$FSR, and (c) 4$\times$FSR. The FSR is 200 GHz. The pump offset of 4$\times$FSR (c) enables the generation of a degenerate signal/idler pair between the pumps, which results in the generation of the bi-phase state.}
\label{Fig5}
\end{figure}

We numerically simulate the degenerate OPO using the modified Lugiato-Lefever equation \cite{Lugiato,Haelterman,Matsko,Coen}, which we previously developed for modeling parametric frequency comb generation in microresonators \cite{Lamont}.  The model includes effects of higher-order dispersion and self-steepening, which enables simulations over a wide range of conditions. The dispersion of the microresonator mode is simulated using a finite-element mode solver [Fig. \ref{Fig4}(a)]. The simulated spectra of the OPO is shown in Fig. \ref{Fig4}(b) for a Si$_3$N$_4$ microresonator with a free spectral range (FSR) of 200 GHz and a cross section of 690$\times$1300 nm, which corresponds to normal GVD at the pump wavelengths of 1535 and 1565 nm for the fundamental TM mode. The two pumps are each offset from the frequency degeneracy point by $\delta=4\times$FSR's. The simulated spectra show that parametric oscillation of the degenerate signal/idler pair occurs only under pump detuning conditions consistent with our analysis above. Multiple simulations with different initial noise conditions indicate that the degenerate signal field settles into one of two phase states which are offset by $\pi$ [Fig. \ref{Fig4}(c)], which is precisely the required behavior for simulating the Ising model as well as for quantum random number generation. Since the generation rate is largely dependent in the cavity lifetime, the current system should be capable of producing random numbers at rates of 100 MHz. We believe that by reducing the size of the resonator and with suitable tuning of the cavity linewidth, random number rates approaching 1 GHz can be readily achieved.

In our experiments, we use a Si$_3$N$_4$ microresonator with an effective resonator cross section of 690$\times$1300 nm, which allows for the necessary normal GVD for the fundamental TM polarization mode. The microresonator has a loaded \emph{Q} of 3 million and an FSR of 200 GHz. We combine the output beams from two amplified tunable single-frequency pump lasers using a 50/50 fiber coupler and inject them into the Si$_3$N$_4$ bus waveguide which is coupled to the microresonator. WDM filters with 9-nm bandwidth are used to remove the amplified spontaneous emission from the erbium-doped fiber amplifier at each of the respective outputs before combining. The input polarization is set to quasi-TM. The output from the device is collected using a free-space objective and sent into a 80/20 fiber coupler such that $20 \%$ of the signal is sent to an optical spectrum analyzer for characterization, and the remaining light is sent to two different WDM filters and a photodiode to monitor independently the transmission of each of the pumps to ensure that they are coupled to a cavity resonance. The combined pump power in the coupling waveguide is 125 mW.  

In order to generate the dual-pumped OPO, we tune each pump into a microresonator cavity resonance while taking into account the thermal frequency shift that occurs as the pump power builds up in the cavity.  The reduction of the pump transmission along with cascaded FWM due to the contribution from the two pumps indicates that the pump is coupled to a cavity resonance. Figure \ref{Fig5} shows the generated OPO spectra for various pump frequency offsets $\delta$ from the degeneracy point of, 6.5$\times$FSR [Fig. \ref{Fig5}(a)], 5$\times$FSR [Fig. \ref{Fig5}(b)], and 4$\times$FSR [Fig. \ref{Fig5}(c)], which corresponds to pump separations of 13$\times$, 10$\times$, and 8$\times$FSR, respectively. Each of the generated spectra shows a gain peak between the two pump wavelengths, consistent with the theoretical results. As discussed earlier, the pump separation along with the pump detuning within the resonance must be carefully chosen to achieve conditions for degenerate oscillation. Figures \ref{Fig5}(a) and (b) show multiple spectral lines generated between the two pumps, with spacings of 1$\times$FSR and 2$\times$FSR, respectively. This spacing depends on the first lines that oscillate near the frequency degeneracy point. Thus, if the degenerate signal/idler is the first spectral line that reaches the oscillation threshold, there will be no other lines generated between the pumps. For the offset of $\delta=4\times$FSR, which corresponds to pump wavelengths of 1545 nm and 1557.7 nm [Fig. \ref{Fig5}(c)], we generate a single frequency component between the two pumps, corresponding to a degenerate signal/idler pair at the center frequency, which will enable the generation of the bi-phase state. 

In conclusion, we theoretically and experimentally demonstrate frequency-degenerate parametric oscillation via dual-pump FWM in Si$_3$N$_4$ microresonators operating in the normal GVD regime. We show that the Si$_3$N$_4$ platform offers the potential of developing a scalable CMOS-compatible platform that utilizes the unique properties of light to solve challenging computational problems and for use as a system for random number generation.
\\
\\
\textbf{Funding.} Defense Advanced Research Projects Agency (QuASAR); Air Force Office of Scientific Research (FA9550-12-1-0377); National Science Foundation (NSF) (ECS-0335765, ECCS-1306035).
\\
 \\
\textbf{Acknowledgment.} This work was performed in part at the Cornell Nano-Scale Facility, a member of the National Nanotechnology Infrastructure Network, which is supported by the NSF. The authors thank Ron Synowicki from J. A. Woollam Co., Inc for the characterization of the Si$_3$N$_4$ refractive index. S.R. is funded by a EU Marie Curie Fellowship (PIOF-GA-2012-329851). We thank A.R. Johnson and C. Joshi for discussions.



\end{document}